\documentclass{Amade}
\usepackage[cp1251]{inputenc}
\usepackage{mathrsfs}
\usepackage[english,russian]{babel}
\usepackage[14pt]{extsizes}
\begin{document}

\clearemptydoublepage

\chapter[GENERALISED DIFFUSION AND WAVE EQUATIONS: RECENT ADVANCES]{GENERALISED DIFFUSION AND WAVE EQUATIONS: RECENT ADVANCES}{T. Sandev$^{1,2}$, R. Metzler$^{3}$ and A.~Chechkin$^{3,4}$ }{$^{1}$Research Center for Computer Science and Information Technologies, \break Macedonian Academy of Sciences and Arts, \break Bul. Krste Misirkov 2, 1000 Skopje, Macedonia \break $^{2}$Institute of Physics, Faculty of Natural Sciences and Mathematics, \break Ss Cyril and Methodius University, Arhimedova 3, 1000 Skopje, Macedonia \break $^{3}$Institute of Physics \& Astronomy, University of Potsdam, \break D-14776 Potsdam-Golm, Germany \break $^{4}$Akhiezer Institute for Theoretical Physics, Kharkov 61108, Ukraine \break e-mail: achechkin@kipt.kharkov.ua}

\authormark{T. SANDEV, R. METZLER, A.~CHECHKIN}

\begin{abstract}
We present a short overview of the recent results in the theory of diffusion and wave equations with generalised derivative operators. We give generic examples of such generalised diffusion and wave equations, which include time-fractional, distributed order, and tempered time-fractional diffusion and wave equations. Such equations exhibit multi-scaling time behaviour, which makes them suitable for the description of different diffusive regimes and characteristic crossover dynamics in complex systems.  
\end{abstract}

\begin{keys}
Memory kernel, mean squared displacement, subordination\\
MSC (2000): 26A33, 33E12
\end{keys}

\section{INTRODUCTION}

Diffusion equations with fractional time and space derivatives instead of the integer ones are widely used to describe anomalous diffusion processes where the mean squared displacement (MSD) scales as a power of time,
\begin{equation}\label{msd intro}
\left\langle x^{2}(t)\right\rangle\simeq t^{\alpha}.
\end{equation} 
Depending on the values of the anomalous diffusion exponent $\alpha$ one distinguishes the cases of subdiffusion for $0<\alpha<1$, normal Brownian diffusion for $\alpha=1$, superdiffusion for $1<\alpha<2$, ballistic motion for $\alpha=2$, and superballistic motion for $\alpha>2$. Well-known examples of anomalous transport include subdiffusion in artificially crowded systems and protein-crowded lipid bilayer membranes \cite{crowded1,crowded2,crowded3}, subiffusive charge carrier motion in semiconductors \cite{semiconductors}, subdiffusive motion of submicron probes in living biological
cells \cite{cells}, superdiffusive tracer motion in chaotic laminar flows \cite{flows}, diffusion in porous inhomogeneous media \cite{porous}, and random search processes \cite{search}, to name but a few.

Modern microscopic techniques such as fluorescence correlation spectroscopy or advanced single particle tracking methods have led to the discovery of a multitude of anomalous diffusion processes in living biological cells and complex fluids, see e.g. the reviews \cite{barkai,hoefling,meroz, metzPCCP,metzCR,saxton} and references therein. With the growing number of anomalous diffusion phenomena it became clear that a wide range of complex systems do not show a unique, mono-scaling behaviour, Eq.~(\ref{msd intro}), but instead demonstrate transitions between different diffusion regimes in the course of time. Such observations put forward the idea that in order to capture the multi-scaling dynamics one may generalise the fractional differential operator in the fractional diffusion equation by more universal operators with specific memory kernels. Here we analyse in detail the different versions of such generalised operators and the specific dynamical crossovers they effect. Special case of a power-law kernel recovers fractional derivative and respectively, the mono-scaling diffusion regime.

\section{GENERALISED DIFFUSION EQUATIONS}

\subsection{Natural and modified forms}

We consider generalised diffusion equations in the so-called natural and modified form as generalizations of the time fractional diffusion equations in the Caputo or Riemann-Liouville sense. 

The generalised diffusion equation in the natural form is given by
\cite{fcaa2015}
\begin{equation}\label{diffusion-like eq memory}
\int_0^t\gamma(t-t')\frac{\partial W(x,t')}{\partial t'}\,dt'=\frac{\partial^2W(x,t)}{\partial x^2},
\end{equation}
where the memory kernel $\gamma(t)$ stands on the left hand side of the equation. We consider zero boundary conditions at infinity, $W(\pm\infty,t)=0$, $\frac{\partial}{\partial x}W(\pm\infty,t)=0$, and initial conditions of the form
\begin{eqnarray}\label{initial_condition diff}
W(x,t=0)= \delta(x).
\end{eqnarray}
In turn, the modified form of the equation is given by
\cite{fcaa2018,csf2017}
\begin{equation}\label{diffusion-like eq memory eta}
\frac{\partial W(x,t)}{\partial t}=\frac{\partial}{\partial t}\int_{0}^{t}\eta(t-t')\frac{\partial^{2}W(x,t')}{\partial x^{2}}\,dt',
\end{equation}
with the memory kernel $\eta(t)$ on the right hand side of the equation. As it was shown in \cite{fcaa2018}, these two equations are simply connected through the memory kernels in the form $\hat{\gamma}(s)\rightarrow1/[s\hat{\eta}(s)]$, where $\hat{\gamma}(s)=\int_{0}^{\infty}e^{-st}\gamma(t)\,dt=\mathscr{L}\left[\gamma(t)\right]$ and $\hat{\eta}(s)=\mathscr{L}\left[\eta(t)\right]$ are the
Laplace transforms of the memory kernels $\gamma(t)$ and $\eta(t)$, respectively. These equations have been obtained from continuous time random walk (CTRW) theory for finite variance of jump lengths and generalised waiting time probability density functions (PDFs) of the forms $\hat{\psi}(s)=\frac{1}{1+s\hat{\gamma}(s)}$ and $\hat{\psi}(s)=\frac{1}{1+[\hat{\eta}(s)]^{-1}}$, respectively.

In order to have well established stochastic processes encoded in both equations, we need to prove that their solutions are normalized and non-negative. The non-negativity of the solutions can be shown by using the subordination approach \cite{cheFCAA,mark,mark2}. We will elaborate on this approach for Eq. (\ref{diffusion-like eq memory}), this can be done for Eq. (\ref{diffusion-like eq memory eta}) in the same way. By Fourier ($\tilde{f}(k)=\int_{-\infty}^{\infty}f(x)e^{\imath k x}\,dx$) and Laplace transformations of Eq. (\ref{diffusion-like eq memory}) one finds
\begin{eqnarray}\label{subordination PDF FL general}
\tilde{\hat{W}}(k,s)=\hat{\gamma}(s)\int_0^{\infty}e^{-u\left(s\hat{\gamma}(s)+k^2\right)}\,du=\int_0^{\infty}e^{-uk^2}\hat{G}(u,s)\,du.
\end{eqnarray}
Here the function $G$ is defined by
\begin{equation}\label{G(u,s) general}
\hat{G}(u,s)=\hat{\gamma}(s)e^{-u\,s\hat{\gamma}(s)}=-\frac{\partial}{\partial u}\frac{1}{s}e^{-u\,s\hat{\gamma}(s)}.
\end{equation}
Therefore, the PDF $W(x,t)$ is given by \cite{mark,mark2}
\begin{eqnarray}\label{PDF diffusion like}
W(x,t)&=&\int_0^{\infty}\frac{e^{-\frac{x^2}{4u}}}{\sqrt{4\pi u}}G(u,t)\,du,
\end{eqnarray}
which means that the function $G(u,t)$ is the PDF providing the subordination transformation from time scale $t$ (physical time) to time scale $u$ (operational time). The function $G(u,t)$ is normalized with respect to $u$ for any $t$, i.e.,
\begin{eqnarray}\label{G(u,s) norm}
\int_{0}^{\infty}G(u,t)\,du=\mathscr{L}_{s}^{-1}\left[\int_{0}^{\infty}\hat{\gamma}(s)e^{-us\hat{\gamma}(s)}\,du\right]=\mathscr{L}_{s}^{-1}\left[\frac{1}{s}\right]=1.\nonumber\\
\end{eqnarray}
In order to prove the positivity of $W(x,t)$ according to the Bernstein theorem it is sufficient to show that the function $\hat{G}(u,s)$ is completely monotone on the positive real axis $s$ \cite{book bernstein}. For that we only need to show that (i) the function $\hat{\gamma}(s)$ is completely monotone, and (ii) the function $s\hat{\gamma}(s)$ is a Bernstein function. If (ii) holds, then the function $e^{-s\hat{\gamma}(s)}$ is completely monotone as a composition of completely monotone and a Bernstein function. Moreover, $G(u,s)$ is completely monotone, as a product of two completely monotone functions, $e^{-s\hat{\gamma}(s)}$ and $\hat{\gamma}(s)$. Alternatively, one can check that $s\hat{\gamma}(s)$ is a complete Bernstein function, which is an important subclass of the Bernstein functions \cite{book bernstein}. This condition is enough to ensure the complete monotonicity of $\hat{G}(u,s)$ due to the property of the complete Bernstein function: if $f(s)$ is a complete Bernstein function, then $f(s)/s$ is completely monotone \cite{book bernstein}. The proof of the non-negativity of the solutions to the generalised diffusion equations with different memory kernels can be found in \cite{fcaa2018,csf2017} along with the list of properties of completely monotone, Bernstein, and complete Bernstein functions. 

By analogy, the solution of Eq.~(\ref{diffusion-like eq memory eta}) is non-negative if (i) the function $1/[s\hat{\eta}(s)]$ is completely monotone, and (ii) the function $1/\hat{\eta}(s)$ is a Bernstein function. Alternatively, one can prove the non-negativity of the solution if $1/\hat{\eta}(s)$ is a complete Bernstein function. 

By solving both equations (\ref{diffusion-like eq memory}) and (\ref{diffusion-like eq memory eta}), one can find the corresponding MSDs, as follows \cite{fcaa2015,fcaa2018,csf2017}
\begin{equation}\label{MSDgeneral n}
\left\langle x^{2}(t)\right\rangle %=\left.\mathcal{L}^{-1}\left[-\frac{\partial^{2}}{\partial k^{2}}W(k,s)\right]\right|_{k=0}
=2\,\mathscr{L}^{-1}\left[\frac{s^{-2}}{\hat{\gamma}(s)}\right],
\end{equation}
\begin{equation}\label{MSDgeneral m}
\left\langle x^{2}(t)\right\rangle=2\,\mathscr{L}^{-1}\left[s^{-1}\hat{\eta}(s)\right],
\end{equation}
from which we can analyze the diffusive behaviours for a given form of the memory kernel.

\subsection{Particular examples}

\subsubsection{2.2.1. Standard diffusion equation}

The case with $\gamma(t)=\delta(t)$, which means $\hat{\eta}(s)=1/[s\hat{\gamma}(s)]=1/s$, i.e., $\eta(t)=1$, leads us both to the standard diffusion equation
\begin{equation}\label{diffusion eq standard}
\frac{\partial W(x,t)}{\partial t}=\frac{\partial^2 W(x,t)}{\partial x^2},
\end{equation}
for Brownian diffusion with linear time dependence of the MSD, $\left\langle x^{2}(t)\right\rangle=2\,\mathscr{L}^{-1}\left[s^{-2}\right]=2\,t$. The waiting time PDF in the corresponding CTRW scheme is exponential, $\psi(t)=\mathscr{L}^{-1}\left[\frac{1}{1+s}\right]=e^{-t}$, which in the long time limit ($s\rightarrow0$) can be used as $\hat{\psi}(s)\simeq1-s$.

\subsubsection{2.2.2. Mono-fractional diffusion equation}

Another well known example is the case of a power-law memory kernel $\gamma(t)=t^{-\alpha}/\Gamma\left(1-\alpha\right)$, $0<\alpha<1$, from where it follows that $\hat{\eta}(s)=s^{-\alpha}$, i.e., $\eta(t)=t^{\alpha-1}/\Gamma(\alpha)$. These memory kernels yield two equivalent formulations of the time fractional diffusion equation, namely,
\begin{equation}\label{diffusion eq caputo form}
{_C}D_{0+}^{\alpha}W(x,t)=\frac{\partial^{2}W(x,t)}{\partial x^{2}},
\end{equation}
and
\begin{equation}\label{diffusion eq rl form}
\frac{\partial W(x,t)}{\partial t}={_{RL}}D_{0+}^{1-\alpha}\frac{\partial^{2}W(x,t)}{\partial x^{2}},
\end{equation}
where 
\begin{align}\label{caputo der}
{_C}D_{t}^{\alpha}f(t)=\frac{1}{\Gamma(n-\alpha)}\int_{0}^{t}\frac{f^{(n)}(t')}{(t-t')^{\alpha+1-n}}\,dt', \quad n-1<\alpha<n,\nonumber\\
\end{align}
and
\begin{align}\label{rl der}
{_{RL}}D_{t}^{\alpha}f(t)=\frac{1}{\Gamma(n-\alpha)}\frac{d^n}{dt^n}\int_{0}^{t}\frac{f(t')}{(t-t')^{\alpha+1-n}}\,dt', \quad n-1<\alpha<n,
\end{align}
are the Caputo and Riemann-Lioville fractional derivatives, respectively \cite{mainardi book}. Both derivatives for $\alpha=n$ become ordinary derivatives, $f^{(n)}(t)$. The corresponding MSD shows subdiffusive behaviour $\left\langle x^{2}(t)\right\rangle=2\,\mathscr{L}^{-1}\left[s^{-\alpha-1}\right]=2\,\frac{t^{\alpha}}{\Gamma(\alpha+1)}$, with the Mittag-Leffler waiting time PDF, $\psi(t)=\mathscr{L}^{-1}\left[\frac{1}{1+s^{\alpha}}\right]=t^{\alpha-1}E_{\alpha,\alpha}\left(-t^{\alpha}\right)$. Here $E_{\alpha,\beta}(-z)=\sum_{n=0}^{\infty}\frac{(-z)^n}{\Gamma(\alpha n+\beta)}$ is the two parameter Mittag-Leffler function \cite{mainardi book}, which has the following asymptotic $E_{\alpha,\beta}(-z)\simeq-\sum_{n=1}^{\infty}\frac{(-z)^{-n}}{\Gamma(\beta-\alpha n)}$ for $z\gg1$. Here we note that in the long time limit the waiting time PDF is of power-law form, $\psi(t)\simeq t^{-1-\alpha}$ \cite{reports}.

\subsubsection{2.2.3. Bi-fractional diffusion equation}

Now we introduce a memory kernel with two power-law functions, $\gamma(t)=B_1\,t^{-\alpha_1}/\Gamma(1-\alpha_1)+B_2\,t^{-\alpha_2}/\Gamma(1-\alpha_2)$, $0<\alpha_1<\alpha_2<1$, $B_1+B_2=1$, which gives rise to the bi-fractional diffusion equation in natural form \cite{chechkin_pre2002},
\begin{align}\label{bi-frac diffusion eq in normal form C}
B_1\,{_{C}}D_{t}^{\alpha_1}W(x,t)+B_2\,{_{C}}D_{t}^{\alpha_2}W(x,t)=\frac{\partial^{2}W(x,t)}{\partial x^{2}}.
\end{align}
Using the relation with the memory kernel $\eta(t)$, $\hat{\eta}(s)=1/[s\hat{\gamma}(s)]=\left[B_1\,s^{\alpha_1}+B_2\,s^{\alpha_2}\right]^{-1}$, we find that $\eta(t)=\frac{1}{B_2}t^{\alpha_2-1}E_{\alpha_2-\alpha_1,\alpha_2}\left(-\frac{B_1}{B_2}t^{\alpha_2-\alpha_1}\right)$, i.e., the equivalent representation to Eq.~(\ref{bi-frac diffusion eq in normal form C}) in the modified form is given by
\begin{align}\label{bi-frac diffusion eq in mod form RL}
\frac{\partial W(x,t)}{\partial t}&=\frac{1}{B_2}\frac{\partial}{\partial t}\int_{0}^{t}(t-t')^{\alpha_2-1}\nonumber\\&\times E_{\alpha_2-\alpha_1,\alpha_2}\left(-\frac{B_1}{B_2}[t-t']^{\alpha_2-\alpha_1}\right)\frac{\partial^{2}W(x,t')}{\partial x^{2}}\,dt'.
\end{align}
The bi-fractional diffusion equation in the natural form is a model to describe \textit{decelerating} subdiffusion, since the MSD is given by \cite{chechkin_pre2002,pre2015}
\begin{align}\label{MSD_double power law eta}
\left\langle x^2(t)\right\rangle=\frac{2t^{\alpha_2}}{B_2}E_{\alpha_2-\alpha_1,\alpha_2+1}
\left(-\frac{B_1}{B_2}t^{\alpha_2-\alpha_1}\right)\simeq\left\lbrace \begin{array}{l l}
2B_2\frac{t^{\alpha_2}}{\Gamma(1+\alpha_2)}, \quad t\ll1,\\
2B_1\frac{t^{\alpha_{1}}}{\Gamma(1+\alpha_{1})}, \quad t\gg1.
\end{array}\right.
\end{align}
The corresponding waiting time PDF is given by \cite{pre2015}
\begin{align}\label{psi two powers}
\nonumber \psi(t)&=\mathscr{L}^{-1}\left[\frac{1}{1+B_1s^{\alpha_1}+B_2s^{\alpha_2}}\right]\\&=\frac{t^{\alpha_2-1}}{B_2}\sum_{n=0}^{\infty}\frac{(-1)^n}{B_2^n}t^{\alpha_2n}E_{\alpha_2-\alpha_1,\alpha_2n+\alpha_2}^{n+1}\left(-\frac{B_1}{B_2}t^{\alpha_2-\alpha_1}\right).
\end{align}
Here $E_{\alpha,\beta}^{\delta}(z)=\sum_{n=0}^{\infty}\frac{(\delta)_{n}}{\Gamma(\alpha n+\beta)}\frac{z^{n}}{n!}$ is the three parameter Mittag-Leffler function \cite{prabhakar}, where $(\delta)_n=\Gamma(\delta+n)/\Gamma(\delta)$ is the Pochhammer symbol. Its asymptotic expansions are given by $E_{\alpha,\beta}^{\delta}(-t^{\alpha})\simeq\frac{1}{\Gamma(\beta)}-\delta\frac{
t^{\alpha}}{\Gamma(\alpha+\beta)}\simeq\frac{1}{\Gamma(\beta)}\exp\left(-\delta
\frac{\Gamma(\beta)}{\Gamma(\alpha+\beta)}t^{\alpha}\right)$ for $t\ll1$, and $E_{\alpha,\beta}^{\delta}(-t^{\alpha})=\frac{t^{-\alpha\delta}}{\Gamma(\delta)}\sum_{n=0}^{\infty}\frac{\Gamma(\delta+n)}{\Gamma(\beta-\alpha(\delta+n))}\frac{(-t^{\alpha})^{-n}}{n!}$ for $0<\alpha<2$ and $t\gg1$ \cite{garrappa,pre2015}.

In accordance to the previous case, we may introduce the bi-fractional diffusion equation in the modified form, where the memory kernel is given by $\eta(t)=B_1\frac{t^{\alpha_1-1}}{\Gamma\left(\alpha_1\right)}+B_2\frac{t^{\alpha_2-1}}{\Gamma\left(\alpha_2\right)}$ with $0<\alpha_1<\alpha_2<1$, $B_{1}+B_{2}=1$, i.e., \cite{chechkin_pre2008,app}
\begin{equation}\label{diffusion-like eq two powers eta}
\frac{\partial W(x,t)}{\partial t} =B_{1}\,{_{RL}}D_{t}^{1-\alpha_{1}}\frac{\partial^2W(x,t)}{\partial x^2}+B_{2}\,{_{RL}}D_{t}^{1-\alpha_{2}}\frac{\partial^2W(x,t)}{\partial x^2}.
\end{equation}
Since $\hat{\gamma}(s)=1/[s\hat{\eta}(s)]=1/[s(B_1\,s^{-\alpha_1}+B_2\,s^{-\alpha_2})]$, we have $\gamma(t)=\frac{1}{B_1}t^{-\alpha_1}E_{\alpha_2-\alpha_1,1-\alpha_1}\left(-\frac{B_2}{B_1}t^{\alpha_2-\alpha_1}\right)$, and the equivalent representation of Eq.~(\ref{diffusion-like eq two powers eta}) in the natural form is given by a Mittag-Leffler memory kernel,
\begin{align}\label{corr to bi frac in normal form C}
\frac{1}{B_1}\int_{0}^{t}(t-t')^{-\alpha_1} &E_{\alpha_2-\alpha_1,1-\alpha_1}\left(-\frac{B_2}{B_1}[t-t']^{\alpha_2-\alpha_1}\right)\nonumber\\&\times\frac{\partial W(x,t')}{\partial t'}\,dt'=\frac{\partial^{2}W(x,t)}{\partial x^{2}}.
\end{align}
The bi-fractional diffusion equation in the modified form is a useful model for the description of \textit{accelerating} diffusion since the MSD is given by \cite{chechkin_pre2008,pre2015}
\begin{align}\label{MSD_double power law eta rl}
\left\langle x^{2}(t)\right\rangle&=2B_{1}t^{\alpha_1}E_{\alpha_2-\alpha_1,\alpha_1+1}^{-1}\left(-\frac{B_2}{B_1}t^{\alpha_2-\alpha_1}\right)\nonumber\\&=2B_{1}\frac{t^{\alpha_1}}{\Gamma(1+\alpha_1)}+2B_{2}\frac{t^{\alpha_2}}{\Gamma(1+\alpha_2)}\simeq\left\lbrace\begin{array}{l l}
2B_{1}\frac{t^{\alpha_1}}{\Gamma(1+\alpha_1)}, \quad t\ll1,\\
2B_{2}\frac{t^{\alpha_2}}{\Gamma(1+\alpha_2)}, \quad t\gg1.
\end{array}\right.
\end{align}

\subsubsection{2.2.4. Tempered time-fractional diffusion equation}

At the end of this section we show two other models that describe transitions from one to another diffusive behaviour. This can be achieved if one introduces an exponential cut-off of the power-law memory kernel of the form $\gamma(t)=e^{-bt}t^{-\alpha}/\Gamma(1-\alpha)$, $0<\alpha<1$, where $b>0$ is the truncation parameter. Therefore, we obtain the tempered fractional diffusion equation in the natural form,
\begin{align}\label{tempered diffusion eq n}
\frac{1}{\Gamma(1-\alpha)}\int_{0}^{t}e^{-b(t-t')}(t-t')^{-\alpha}\frac{\partial W(x,t')}{\partial t'}\,dt'=\frac{\partial^{2}W(x,t)}{\partial x^{2}}.
\end{align} 
The corresponding equation in the modified form reads
\begin{align}\label{bi-frac diffusion eq in mod form C}
\frac{\partial W(x,t)}{\partial t}=\frac{\partial}{\partial t}\int_{0}^{t}(t-t')^{\alpha-1}E_{1,\alpha}^{-(1-\alpha)}\left(-b[t-t']\right)\frac{\partial^{2}W(x,t')}{\partial x^{2}}\,dt',
\end{align}
since $\eta(t)=\mathscr{L}^{-1}\left[\frac{s^{-1}}{(s+b)^{\alpha-1}}\right]=t^{\alpha-1}E_{1,\alpha}^{\alpha-1}(-bt)$. Equation (\ref{bi-frac diffusion eq in mod form C}) is actually the diffusion equation
\begin{align}\label{bi-frac diffusion eq in normal form C2}
\frac{\partial W(x,t)}{\partial t}={_{RL}}\mathcal{D}_{1,-b,0+}^{1-\alpha,1-\alpha}\frac{\partial^{2}W(x,t)}{\partial x^{2}}.
\end{align}
with the Prabhakar derivative ($0<\mu<1$, $\rho>0$) \cite{garra}
\begin{align}\label{prabhakar nonregularized}
{_{RL}}\mathcal{D}^{\gamma,\mu}_{\rho,\omega,0+}f(x)=\frac{d}{dt}\int_0^{t}(t-t')^{-\mu}E^{-\gamma}_{\rho,1-\mu}\left(\omega[t-t']^{\rho}\right)f(t')\,dt'.
\end{align}
The MSD shows a crossover from subdiffusion to normal diffusion,
\begin{align}\label{MSD_truncated power law}
\left\langle x^{2}(t)\right\rangle=2t^{\alpha}E_{1,\alpha+1}^{\alpha-1}\left(-bt\right)\simeq\left\lbrace\begin{array}{l l}
2\frac{t^{\alpha}}{\Gamma(1+\alpha)}, \quad t\ll1,\\
2b^{1-\alpha}t, \quad t\gg1.
\end{array}\right.
\end{align}

Consider now the tempered fractional diffusion equation in the modified form with $\eta(t)=e^{-bt}t^{\alpha-1}/\Gamma(\alpha)$ ($0<\alpha<1$, $b>0$), i.e.,
\begin{align}\label{mod temp diff eq}
\frac{\partial W(x,t)}{\partial t}=\frac{1}{\Gamma(\alpha)}\frac{\partial}{\partial t}\int_{0}^{t}e^{-b(t-t')}(t-t')^{\alpha-1}\frac{\partial^2 W(x,t')}{\partial x^2}\,dt'.
\end{align}
From $\hat{\eta}(s)=(s+b)^{-\alpha}$ it follows that $\hat{\gamma}(s)=1/[s(s+b)^{-\alpha}]$, and 
$\gamma(t)=t^{-\alpha}E_{1,1-\alpha}^{-\alpha}\left(-bt\right)$, i.e., the corresponding equation in the natural form equivalent to (\ref{mod temp diff eq}) is given by
\begin{align}
\int_{0}^{t}(t-t')^{-\alpha}E_{1,1-\alpha}^{-\alpha}\left(-b[t-t']\right)\frac{\partial W(x,t')}{\partial t'}\,dt'=\frac{\partial^{2}W(x,t)}{\partial x^{2}},
\end{align}
which can be presented with the regularized Prabhakar derivative \cite{garra}
\begin{align}\label{prabhakar regularized}
{_{C}}\mathcal{D}^{\gamma,\mu}_{\rho,\omega,0+}f(x)=\int_0^{t}(t-t')^{-\mu}E^{-\gamma}_{\rho,1-\mu}\left(\omega[t-t']^{\rho}\right)\frac{df(t')}{dt'}\,dt',
\end{align}
as
\begin{align}
{_C}\mathcal{D}_{1,-b,0+}^{\alpha,\alpha}W(x,t)=\frac{\partial^{2}W(x,t)}{\partial x^{2}}.
\end{align}
The corresponding MSD shows a crossover from subdiffusion to a plateau value,
\begin{align}\label{MSD_truncated power law m}
\left\langle x^{2}(t)\right\rangle=2t^{\alpha}E_{1,\alpha+1}^{\alpha+1}\left(-bt\right)\simeq\left\lbrace\begin{array}{l l}
2\frac{t^{\alpha}}{\Gamma(1+\alpha)}, \quad t\ll1,\\
2b^{-\alpha}, \quad t\gg1.
\end{array}\right.
\end{align}

Here we note that different models based on the tempered versions of the generalised Langevin equation and fractional Brownian motion have been introduced recently, which also give similar crossovers from subdiffusion to normal diffusion \cite{njp}. Moreover, general diffusion equations on two dimensional structures have been analyzed and different diffusive regimes obtained \cite{mmnp}.

\section{Generalised diffusion-wave equation}

\subsection{Natural and modified forms}

In analogy to the generalised diffusion equations in natural and modified forms, we now consider the generalised diffusion-wave equation
\begin{eqnarray}\label{distributed order wave eq memory}
\int_{0}^{t}\zeta(t-t')\frac{\partial^{2}W(x,t)}{\partial t'^{2}}\,dt'=\frac{\partial^{2}W(x,t)}{\partial x^{2}},
\end{eqnarray}
in the natural form with non-negative memory kernel $\zeta(t)$, and similarly
%Here we note that by analogy of the generalized diffusion equation in modified form one can introduce the generalized diffusion-wave equation in modified form
\begin{eqnarray}\label{distributed order wave eq memory mod}
\frac{\partial^{2}W(x,t)}{\partial t^{2}}=\frac{\partial^2}{\partial t^2}\int_{0}^{t}\xi(t-t')\frac{\partial^{2}W(x,t')}{\partial x^{2}}\,dt'
\end{eqnarray}
in the modified form with non-negative memory kernel $\xi(t)$. In what follows we consider the natural form, only \cite{jpa2019}. 
%which analysis we omit here, but one can perform all the calculations as those for the equation in natural form. 
The boundary conditions at infinity are $W(\pm\infty,t)=0$, $\frac{\partial}{\partial x}W(\pm\infty,t)=0$, and the initial conditions are of the form
\begin{eqnarray}\label{initial_condition}
W(x,t=0)= \delta(x), \quad \frac{\partial}{\partial t}W(x,t=0)= 0.
\end{eqnarray}
We here refer to \cite{jpa2019} for discussion on the choice of the initial conditions. 

%If we want to consider the solution $W(x,t)$ of  to be PDF then we should find the restrictions on the memory kernel $\zeta(t)$ for which the solution is non-negative. 
Making the Fourier-Laplace transform of Eq.~(\ref{distributed order wave eq memory}), and then inverse Fourier transform, we find
\begin{align}\label{PDF L}
\hat{W}(x,s)=\frac{1}{2}\sqrt{\hat{\zeta}(s)}\exp\left(-s\sqrt{\hat{\zeta}(s)}|x|\right).
\end{align}
From here one easily concludes that the PDF is normalized to 1, i.e., $\int_{-\infty}^{\infty}W(x,t)\,dx=1$, since $\int_{-\infty}^{\infty}\hat{W}(x,s)\,dx=1/s$. The non-negativity of the solution can be shown by applying the Bernstein theorem, i.e., by showing that the solution in the Laplace space is a completely monotone function \cite{book bernstein}. To this end, solution (\ref{PDF L}) can be considered as a product of two functions, $\frac{1}{2}\sqrt{\hat{\zeta}(s)}$ and $\exp\left(-s\sqrt{\hat{\zeta}(s)}|x|\right)$, and it is sufficient to prove that both functions $\sqrt{\hat{\zeta}(s)}$ and $\exp\left(-s\sqrt{\hat{\zeta}(s)}|x|\right)$ are completely monotone. Therefore, it is sufficient to show that $\sqrt{\hat{\zeta}(s)}$ is completely monotone, and $s\sqrt{\hat{\zeta}(s)}$ is a Bernstein function. The non-negativity of the solution can also be shown by proving that the function $\sqrt{\hat{\zeta}(s)}$ is a Stieltjes function, which is again completely monotone, or that $s\sqrt{\hat{\zeta}(s)}$ is a complete Bernstein function. The proof of the non-negativity of the solutions of the generalised diffusion-wave equations with different memory kernels can be found in \cite{jpa2019} along with the list of properties of completely monotone, Stieltjes, Bernstein, and complete Bernstein functions. 

By solving Eq.~(\ref{distributed order wave eq memory}) we find the MSD,
\begin{align}\label{second moment}
\left\langle
x^{2}(t)\right\rangle=\left.\left\{-\frac{\partial^{2}}{\partial k^{2}}\mathscr{L}^{-1}\left[\tilde{\hat{W}}(k,s)\right](k,t)\right\}\right|_{k=0}=2\,\mathscr{L}^{-1}\left[\frac{1}{s^{3}\hat{\zeta}(s)}\right](t),
\end{align}
from where we analyze the diffusive regimes depending on the memory kernel $\zeta(t)$.

\subsection{Particular cases}

\subsubsection{3.2.1. Standard wave equation}

The simplest case of Eq.~(\ref{distributed order wave eq memory}) is the one with Dirac delta memory kernel $\zeta(t)=\delta(t)$, which yields the classical wave equation
\begin{align}\label{classical cattaneo eq}
\frac{\partial^{2}W(x,t)}{\partial t^{2}}=\frac{\partial^{2}W(x,t)}{\partial x^{2}}.
\end{align}
The MSD~(\ref{second moment}) then becomes
\begin{eqnarray}\label{classical cattaneo eq MSD}
\left\langle x^{2}(t)\right\rangle=t^{2},
\end{eqnarray}
which reports ballistic motion.

\subsubsection{3.2.2. Mono-fractional diffusion-wave equation}

The case with the power-law memory kernel $\zeta(t)=\frac{t^{1-\alpha}}{\Gamma(2-\alpha)}$, $0<\alpha<2$, yields
\begin{align}\label{frac diff mu>1/2}
{_{C}D_{t}^{\alpha}}W(x,t)=\frac{\partial^{2}W(x,t)}{\partial x^{2}},
\end{align}
for $1<\alpha<2$, whereas for the case with $0<\alpha<1$ we get
\begin{align}\label{frac diff mu<1/2}
\frac{1}{\Gamma(2-\alpha)}\int_{0}^{t}(t-t')^{1-\alpha}\frac{\partial^{2}W(x,t')}{\partial t'^{2}}\,dt'=\frac{\partial^{2}W(x,t)}{\partial x^{2}}. 
\end{align}
The MSD for the mono-fractional diffusion-wave equation reads 
\begin{eqnarray}\label{MSD 1}
\left\langle x^{2}(t)\right\rangle=2\,\frac{t^{\alpha}}{\Gamma(1+\alpha)}. 
\end{eqnarray}
Since $0<\alpha<2$, the generalised diffusion-wave equation (\ref{distributed order wave eq memory}) with power-law memory kernel describes both superdiffusive and subdiffusive processes. The case $\alpha=1$ reduces to the classical diffusion equation for Brownian motion, i.e., $\left\langle x^{2}(t)\right\rangle=2\,t$, whereas the case with $\alpha=2$ yields ballistic diffusion, $\left\langle x^{2}(t)\right\rangle=t^{2}$.

\subsubsection{3.2.3. Bi-fractional diffusion-wave equation}

The next case we consider is the bi-fractional diffusion-wave equation with the memory kernel of the form $\eta(t)=B_{1}\frac{t^{1-\alpha_{1}}}{\Gamma(2-\alpha_{1})}+B_{2}\frac{t^{1-\alpha_{2}}}{\Gamma(2-\alpha_{2})}$, $B_{1}+B_{2}=1$. The case with $1<\alpha_1<\alpha_2<2$ yields
\begin{align}\label{two power law}
B_{1}\,{_C}D_{t}^{\alpha_{1}}W(x,t)+B_{2}\,{_C}D_{t}^{\alpha_{2}}W(x,t)=\frac{\partial^{2}W(x,t)}{\partial x^{2}},
\end{align}
where ${_C}D_{t}^{\alpha_{j}}$ is the Caputo fractional derivative (\ref{caputo der}) of the order $1<\alpha_{j}<2$ ($n=2$), whereas the case $0<\alpha_{1}<\alpha_{2}<1$ yields equation
\begin{align}\label{distributed order wave eq memory p 2 delta}
&\frac{B_{1}}{\Gamma(2-\alpha_{1})}\int_{0}^{t}(t-t')^{1-\alpha_{1}}\frac{\partial^{2}W(x,t')}{\partial t'^{2}}\,dt'\nonumber\\&+\frac{B_{2}}{\Gamma(2-\alpha_{2})}\int_{0}^{t}(t-t')^{1-\alpha_{2}}\frac{\partial^{2}W(x,t')}{\partial t'^{2}}\,dt'=\frac{\partial^{2}W(x,t)}{\partial x^{2}}.
\end{align}
The corresponding MSD then becomes
\begin{align}\label{msdPower2}
\left\langle x^{2}(t)\right\rangle&=\frac{2}{B_{2}}t^{\alpha_{2}}E_{\alpha_2-\alpha_1,\alpha_2+1}\left(-\frac{B_{1}}{B_{2}}t^{\alpha_{2}-\alpha_{1}}\right),\nonumber\\&\simeq\left\lbrace\begin{array}{l l}
\frac{2}{B_2}\frac{t^{\alpha_2}}{\Gamma(1+\alpha_2)}, \quad & t\ll1 ,\\
\frac{2}{B_1}\frac{t^{\alpha_1}}{\Gamma(1+\alpha_1)}, \quad & t\gg1 ,
\end{array}\right. 
\end{align}
which means {\it decelerating superdiffusion} for $1<\alpha_{1}<\alpha_{2}<2$, including crossover from superdiffusion to normal diffusion in the case $1=\alpha_1<\alpha_2<2$, and {\it decelerating subdiffusion} for $0<\alpha_{1}<\alpha_{2}<1$, including crossover from normal diffusion to subdiffusion for the case $0<\alpha_1<\alpha_2=1$. Decelerating superdiffusion has indeed been observed, for example, in Hamiltonian systems with long-range interactions \cite{latora}, and different biological systems \cite{caspi}.

\subsubsection{3.2.4. Tempered time-fractional wave equation}

Furthermore, we consider a truncated power-law memory kernel of the form
$\zeta(t)=e^{-bt}\frac{t^{1-\alpha}}{\Gamma(2-\alpha)}$, where $b>0$, and $1\le\alpha<2$, corresponding to the following tempered fractional wave equation:
\begin{align}\label{tempered fractional wave eq}
\frac{1}{\Gamma(2-\alpha)}\int_{0}^{t}e^{-b(t-t')}(t-t')^{1-\alpha}\frac{\partial^{2}W(x,t)}{\partial t'^{2}}\,dt'=\frac{\partial^{2}W(x,t)}{\partial x^{2}}.
\end{align}
For the MSD we get
\begin{align}
\left\langle x^{2}(t)\right\rangle=2\,{_{RL}}I_{t}^{3}\left(e^{-bt}\frac{t^{-3+\alpha}}{\Gamma(-2+\mu)}\right)\simeq\left\lbrace\begin{array}{l l}
2\,\frac{t^{\alpha}}{\Gamma(1+\alpha)}, \quad & t\ll1,\\
b^{2-\alpha}t^{2}, \quad & t\gg1,
\end{array}\right. 
\end{align}
where
\begin{align}
{_{RL}I_t^{\alpha}}f(t)=\frac{1}{\Gamma(\alpha)}\int_0^t(t-t')^{\alpha-1}f(t')\,dt', \quad \alpha>0,
\end{align} 
is the Riemann-Liouville integral \cite{mainardi book}. Thus, there is a crossover from superdiffusion to ballistic motion in the case with $1<\alpha<2$, and from normal diffusion to ballistic motion in the case with $\alpha=1$. For the case of the diffusion-wave equation with Prabhakar derivative we address the reader to \cite{jpa2019}.

\section{Summary}

We consider different stochastic processes governed by the generalised diffusion and diffusion-wave equations which contain the well known time fractional diffusion and wave equations as particular cases. Such processes demonstrate a rich multi-scaling behaviour which manifests itself in specific crossovers between different diffusion regimes in the course of time. We thus obtain a flexible tool which can be applied for the description of diverse diffusion phenomena in complex systems demonstrating crossover behaviours.

\vspace{3mm}

\section*{ACKNOWLEDGEMENTS}
The Authors acknowledge funding from the Deutsche Forschungsge\-meinschaft (DFG), project ME 1535/6-1 "Random search processes, L\'evy flights, and random walks on complex networks". RM thanks the Foundation for Polish Science for support within an Alexander von Humboldt Polish Honorary Research Scholarship. RM and AC also acknowledge support from the DFG project 1535/7-1 "Mathematical and physical modeling of single particle tracking - Big Data approach".

%\section*{REFERENCES}


\begin{thebibliography}{99}

\bibitem{barkai} 
Barkai E., Garini Y., Metzler R. Strange kinetics of single molecules in living cells, {\it Phys. Today} {\bf65} (2012), 29--35.


\bibitem{caspi}
Caspi A., Granek R. Elbaum M. Enhanced diffusion in active intracellular transport, {\it Phys. Rev. Lett.} {\bf85} (2000), 5655.

\bibitem{chechkin_pre2002}
Chechkin A.V., Gorenflo R., Sokolov I.M. Retarding subdiffusion and accelerating superdiffusion governed by distributed-order fractional diffusion equations, {\it Phys. Rev. E} {\bf66} (2002), 046129.

\bibitem{cheFCAA} 
Chechkin A.V., Gorenflo R., Sokolov I.M., Gonchar V.Yu. Distributed order time fractional diffusion equation. {\it Fract. Calc. Appl. Anal.} {\bf6} (2003), 259--279.

\bibitem{chechkin_pre2008}
Chechkin A.V., Gonchar V.Yu., Gorenflo R., Korabel N., Sokolov I.M. Generalized fractional diffusion equations for accelerating subdiffusion and truncated L\'evy flights, {\it Phys. Rev. E} {\bf78} (2008), 021111.

\bibitem{garrappa}
Garra R., Garrappa R. The Prabhakar or three parameter Mittag--Leffler function: Theory and application, {\it Commun. Nonlinear Sci. Numer. Simul.} {\bf56} (2018), 314--329.

\bibitem{garra}
Garra R., Gorenflo R., Polito F., Tomovski Z. Hilfer--Prabhakar derivatives and some applications, {\it Appl. Math. Comput.} {\bf242} (2014), 576--589.

\bibitem{cells}
Golding I., Cox E.C. Physical nature of bacterial cytoplasm, {\it Phys. Rev. Lett.} {\bf96} (2006), 098102.

\bibitem{hoefling} 
Hoefling F., Franosch T. Anomalous transport in the crowded world of biological cells, {\it Rep. Prog. Phys.} {\bf76} (2013), 046602.

\bibitem{crowded1}
Jeon J.-H., Leijnse N., Oddershede L.B., Metzler R. Anomalous diffusion and power-law relaxation of the time averaged mean squared displacement in worm-like micellar solutions, {\it New J. Phys.} {\bf15} (2013), 045011.

\bibitem{crowded2}
Jeon J.-H., Monne H.M.-S., Javanainen M., Metzler R. Anomalous diffusion of phospholipids and cholesterols in a lipid bilayer and its origins, {\it Phys. Rev. Lett.} {\bf109} (2012), 188103.

\bibitem{latora}
Latora V., Rapisarda A., Ruffo S. Superdiffusion and out-of-equilibrium chaotic dynamics with many degrees of freedoms, {\it Phys. Rev. Lett.} {\bf83} (1999), 2104.


\bibitem{mainardi book}
Mainardi F. {\it Fractional Calculus and Waves in Linear Viscoelesticity:  An introduction to Mathematical Models}, Imperial College Press,  London, 2010.


\bibitem{mark}
Meerschaert M.M., Benson D.A., Scheffler H.P., Baeumer B. Stochastic solution of space-time fractional diffusion equations, {\it Phys. Rev. E} {\bf65} (2002), 041103. 

\bibitem{mark2}
Meerschaert M.M., Straka P. Inverse stable subordinators, {\it Math. Model. Nat. Phenom.} {\bf8} (2013), 1--16.

\bibitem{meroz} 
Meroz Y., Sokolov I.M. A toolbox for determining subdiffusive mechanisms, {\it Phys. Rep.} {\bf573} (2015), 1--29.

\bibitem{reports}
Metzler R., Klafter J. The random walk's guide to anomalous diffusion: a fractional dynamics approach, {\it Phys. Rep.} {\bf339} (2000), 1--77.

\bibitem{metzPCCP} 
Metzler R., Jeon J.-H., Cherstvy A.G., Barkai E. Anomalous diffusion models and their properties: non-stationarity, non-ergodicity, and ageing at the centenary of single particle tracking, {\it Phys. Chem. Chem. Phys.} {\bf16} (2014), 24128--24164.

\bibitem{metzCR} 
N{\o}rregaard K., Metzler R., Ritter C.M., Berg-S{\o}rensen K., Oddershede L.B. Manipulation and motion of organelles and single molecules in living cells, {\it Chem. Rev.} {\bf117} (2017), 4342--4375.

\bibitem{njp}
Molina-Garcia D., Sandev T., Safdari H., Pagnini G., Chechkin A., Metzler R. Crossover from anomalous to normal diffusion: truncated power-law noise correlations and applications to dynamics in lipid bilayers, {\it New J. Phys.} {\bf20} (2018), 103027.

\bibitem{prabhakar}
Prabhakar T.R. A singular integral equation with a generalized Mittag Leffler function in the kernel, {\it Yokohama Math. J.} {\bf19} (1971), 7--15.

\bibitem{fcaa2015}
Sandev T., Chechkin A., Kantz H., Metzler R. Diffusion and Fokker-Planck-Smoluchowski equations with generalized memory kernel, {\it Fract. Calc. Appl. Anal.} {\bf18} (2015), 1006--1038.

\bibitem{pre2015}
Sandev T., Chechkin A.V., Korabel N., Kantz H., Sokolov I.M., Metzler R. Distributed-order diffusion equations and multifractality: Models and solutions, {\it Phys. Rev. E} {\bf92} (2015), 042117.

\bibitem{mmnp}
Sandev T., Iomin A., Kantz H., Metzler R., Chechkin A. Comb model with slow and ultraslow diffusion, {\it Math. Model. Nat. Phenom.} {\bf11} (2016), 18--33.

\bibitem{fcaa2018}
Sandev T., Metzler R., Chechkin A. From continuous time random walks to the generalized diffusion equation, {\it Fract. Calc. Appl. Anal.} {\bf21} (2018), 10--28.

\bibitem{csf2017}
Sandev T., Sokolov I.M., Metzler R., Chechkin A. Beyond monofractional kinetics, {\it Chaos, Solitons \& Fractals} {\bf102} (2017), 210--217.

\bibitem{jpa2019}
Sandev T., Tomovski Z., Dubbeldam J.L.A., Chechkin A. Generalized diffusion-wave equation with memory kernel, {\it J. Phys. A: Math. Theor.} {\bf52} (2019), 015201.

\bibitem{saxton} 
Saxton M.J., Jacobsen K. Single-particle tracking: applications to membrane dynamics, {\it Annu. Rev. Biophys. Biomol. Struct.} {\bf26} (1997), 373--399.

\bibitem{semiconductors}
Scher H., Montroll E.W. Anomalous transit-time dispersion in amorphous solids, {\it Phys. Rev. B} {\bf12} (1975), 2455.

\bibitem{book bernstein}
Schilling R., Song R., Vondracek Z. {\it Bernstein Functions}, De Gruyter, Berlin, 2010.

\bibitem{app}
Sokolov I.M., Chechkin A.V., Klafter J. Distributed-order fractional kinetics, {\it Acta Phys. Pol. B} {\bf35} (2004), 1323--1341.

\bibitem{flows}
Solomon T.H., Weeks E.R., Swinney H.L. Observation of anomalous diffusion and L\'evy flights in a two-dimensional rotating flow, {\it Phys. Rev. Lett.} {\bf71} (1993), 3975.

\bibitem{crowded3}
Szymanski J., Weiss M. Elucidating the origin of anomalous diffusion in crowded fluids, {\it Phys. Rev. Lett.} {\bf103} (2009), 038102.

\bibitem{search}
Viswanathan G.E., da Luz M.G.E., Raposo E.P., Stanley H.E. {\it The Physics of Foraging. An Introduction to Random Searches and Biological Encounters}, Cambridge University Press, Cambridge, 2011.

\bibitem{porous}
Zhokh A., Trypolskyi A., Strizhak P. Relationship between the anomalous diffusion and the fractal dimension of the environment, {\it Chem. Phys.} {\bf503} (2018), 71--76.

\end{thebibliography}
\end{document}